\documentclass{article}

\usepackage[T1]{fontenc}
\usepackage{microtype,graphicx,xcolor,xspace}
\usepackage{amsmath,amssymb}

\usepackage{array,tabularx,longtable,booktabs,colortbl}
\usepackage{float,needspace}

\usepackage{algorithm,algorithmic}
\usepackage{xurl,hyperref}

\usepackage[accepted]{mlsys2025}

\hypersetup{
  pdfauthor={Rahul Sharma, Nitin Vegesna, Yanxin Zhang},
  pdfsubject={Preprint},
  pdftitle={Energy-Efficient Reinforcement Learning for Large Language Models}
}

\newcommand{\system}{EE-RL\xspace}
\newcommand{\paperkeywords}{LLM post-training, reinforcement learning, GPU energy efficiency, dynamic voltage and frequency scaling (DVFS), runtime frequency control, phase-aware optimization, asynchronous reinforcement learning, energy-aware scheduling}
\mlsystitlerunning{Energy-Efficient Reinforcement Learning for Large Language Models}

\def\hdashline[#1]{%
  \noalign{%
    \vskip2pt
    \hbox to\linewidth{%
      \color{black!35}%
      \leaders\hbox{\kern1pt\rule{0.5pt}{0.5pt}\kern1pt}\hfill
    }%
    \vskip1pt
  }%
}
\begin{document}
\twocolumn[
\mlsystitle{Energy-Efficient Reinforcement Learning\texorpdfstring{\\}{ }for Large Language Models}
\mlsyssetsymbol{equal}{*}
\mlsyssetsymbol{corresponding}{\ensuremath{\dagger}}
\begin{mlsysauthorlist}
\mlsysauthor{Rahul Sharma}{equal}
\mlsysauthor{Nitin Vegesna}{equal}
\mlsysauthor{Yanxin Zhang}{corresponding}
\par\vspace{0.4em}
{\normalfont NVIDIA}
\end{mlsysauthorlist}
\mlsyskeywords{\paperkeywords}
\vskip 0.3in
\begin{abstract}
Reinforcement learning for language models alternates between generating responses and training the policy. These stages place different demands on GPUs, creating opportunities to reduce clock frequency with little loss in throughput. Selecting frequencies online is difficult because response lengths change and frequency trials themselves consume time and energy. We present \system (energy-efficient RL), an online controller that selects separate frequencies for generation and training during RL execution. It compares candidate settings with a prediction of the default GPU governor on the same token workload, and keeps the most energy-efficient setting that meets a throughput threshold. The controller then monitors throughput and energy and adjusts frequencies as the workload changes. Fixed-frequency sweeps on H100 and GB200 show GPU energy savings of up to 28.2\% per valid training token across synchronous and asynchronous execution, while largely preserving throughput. \system turns this headroom into online savings, reducing GPU energy per valid training token by up to 16.9\% at a modest throughput cost.
\par\smallskip
{\small\noindent\textbf{Keywords:} \paperkeywords.\par}
\end{abstract}
]
\begingroup
\renewcommand{\thefootnote}{}
\makeatletter
\renewcommand{\@makefntext}[1]{\setlength{\parindent}{0pt}\noindent#1}
\makeatother
\footnotetext{\makebox[0pt][r]{\textsuperscript{*}\,}Equal contribution; authors listed in random order.\newline
Rahul Sharma: \href{mailto:rahulsha@nvidia.com}{rahulsha@nvidia.com}.\newline
Nitin Vegesna: \href{mailto:nvegesna@nvidia.com}{nvegesna@nvidia.com}.\newline
\makebox[0pt][r]{\textsuperscript{\ensuremath{\dagger}}\,}Corresponding author: Yanxin Zhang
(\href{mailto:yanxinz@nvidia.com}{yanxinz@nvidia.com}).
\par\smallskip\noindent Preprint.}
\endgroup
\section{Introduction}
\label{sec:introduction}
Reinforcement learning (RL) post-training improves a language model by repeatedly generating responses, evaluating them, and updating the model~\cite{sheng2024hybridflow,fu2025areal}. This process can consume substantial energy. Post-training Olmo 3 32B as a reasoning model, for example, used 17$\times$ more datacenter energy than its instruction-tuned counterpart, largely due to rollout generation~\cite{morrison2026hidden}. The growing electricity demand of AI data centers makes energy efficiency important for operating costs and infrastructure expansion~\cite{iea2025energyai}. Reducing energy consumption without substantially extending GPU allocation time can lower electricity costs while preserving resource availability. Because GPUs consume energy while computing, moving data, and waiting, improving energy efficiency requires identifying when higher clock frequencies accelerate post-training and when they merely increase power consumption.

\begin{figure}[!t]
\centering\includegraphics[width=\columnwidth]{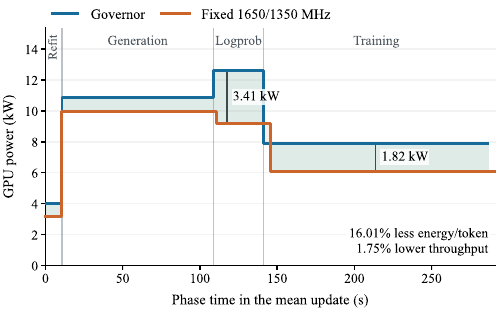}
\caption{Phase-average GPU power in Qwen3-30B-A3B on 32 H100s under the governor and at fixed 1650/1350\,MHz generation/training frequencies. The time axis omits inter-phase gaps, which the annotated changes include.}
\label{fig:governor-profile}
\end{figure}

An RL update includes response generation, log-probability computation, policy training, and weight transfer to the generation engine. These phases differ in arithmetic intensity~\cite{williams2009roofline} and therefore place different demands on the hardware. Autoregressive generation produces responses one token at a time and can be limited by memory bandwidth, while training processes batches of complete responses and typically uses more parallel computation. Weight transfers, long responses that leave the rest of a batch waiting~\cite{rollpacker2025}, and dependencies between phases introduce more waiting. Default GPU management, which we call the \emph{governor}, nevertheless runs all of this work near peak clock frequency, so memory-bound work and idle waiting can consume additional power without a corresponding throughput gain.

This mismatch creates \emph{headroom}, an opportunity to lower GPU frequency, and with it power, while losing little end-to-end throughput relative to the governor. The benefit depends on whether lower power outweighs any increase in execution time. Figure~\ref{fig:governor-profile} illustrates the opportunity by comparing power under the governor with a lower-frequency setting. Power is lower in all phases, while their durations remain similar. The available headroom varies with the workload and how generation and training are distributed across GPUs.

Finding an efficient setting in advance is expensive. It requires repeated RL runs that consume valuable GPU hours, and the result may change with the model, batch size, or hardware allocation. A setting chosen at startup can also become less suitable as the model produces longer or shorter responses~\cite{rollpacker2025,laminar2025}. Selecting frequencies online avoids a separate search for every workload, but the controller must distinguish these changes in work from the effect of changing frequency. It must also save enough energy to offset the cost of testing candidate settings.

We present \system (energy-efficient RL), a controller that selects separate generation and training frequencies while the RL workload runs. It measures GPU energy per valid training token and throughput in valid training tokens per second, rather than per step, because response lengths vary and steps do not represent a fixed amount of work. \emph{Valid training tokens} are response tokens consumed by training that contribute to the loss. The controller tests frequency settings in short trials against the governor. It keeps the most energy-efficient setting within its throughput limit, then monitors energy and throughput to adjust frequencies as needed.

The paper makes three contributions:
\begin{itemize}
\item[$\bullet$] \textbf{Evidence of energy-saving headroom.} Measurements on H100 and GB200 show where lower phase frequencies save energy and where longer execution offsets the power reduction.
\item[$\bullet$] \textbf{Online per-phase frequency control.} \system accounts for changing token workloads and selects the most energy-efficient tested setting that satisfies its throughput bound relative to the governor.
\item[$\bullet$] \textbf{Controller evaluation.} Matched comparisons measure energy savings and throughput over the full measured workload in synchronous and asynchronous RL, including the cost of testing and changing frequencies.
\end{itemize}

\section{Background: How RL Uses GPU Resources}
\label{sec:background}

Higher GPU clock frequencies can speed up compute-intensive work, but offer less benefit when performance is limited by memory movement or waiting. RL includes phases with different demands, and the GPU allocation also influences their execution time and compute demands. These differences create opportunities to lower frequencies---and energy consumption---with little loss in overall performance.

\subsection{Generation, training, and useful work}
Our workloads use Group Relative Policy Optimization (GRPO), which generates several responses to each prompt and compares their rewards within the group to update the model~\cite{shao2024deepseekmath}. We call this repeated process an update. Generation produces the responses, and training uses them to change the model's parameters. The updated parameters are then made available to the generation engine.

Each update also includes log-probability computation and weight refit. Log-probability computation scores response tokens for the training objective, while weight refit transfers updated policy parameters to the generation engine. These stages can have different power demands, but their shorter durations make separate frequency settings less worthwhile. To limit clock-switching overhead, we set log-probability computation to use the training frequency, while weight refit runs under both. Energy measurements include all allocated GPUs, covering generation, training, weight transfer, and waiting.

\subsection{How generation and training share GPUs}
We consider three execution modes. In \textbf{colocated synchronous} execution, generation and training take turns on the same GPUs. Both contribute directly to the time required for an update, and frequency changes follow their phase boundaries.

In \textbf{disaggregated synchronous} execution, generation and training use separate GPU pools but remain synchronized at each update. Training waits for the generated batch, and the next generation round waits for updated policy weights. Pool sizes can differ, and the waiting pool continues to consume power.

In both synchronous modes, frequency changes preserve the intended update dependencies. Each training update consumes the completed batch before the next generation round uses the updated policy. Unlike asynchronous execution, changing a phase's duration does not change which batch is scheduled for that update, so frequency changes should not affect the model's final accuracy.

In \textbf{asynchronous} execution, generation produces responses while training consumes previously generated work. A faster pool may therefore tolerate a lower frequency without slowing overall post-training progress. Timing can also affect which responses are available when training samples a batch. We define a response's \emph{policy age} as the difference between the policy version used for generation and the version against which it is consumed for training. Bounding this age limits the staleness of consumed responses. 

We evaluate two weight-update schedules. With \emph{drained} updates, in-progress responses finish before new weights are loaded into the generation engine. With \emph{in-flight} updates, generation pauses with responses still in progress, loads new weights, recomputes their KV caches, and resumes. The policy-age limits used by each schedule are reported with the results.

\subsection{Power, time, and energy}
The governor serves as our reference, with observed peak clocks of 1980\,MHz on H100 and 2062\,MHz on GB200. \system requests GPU core frequencies while hardware manages voltage. Dynamic power scales approximately as $P_{\mathrm{dyn}}\propto V_{\mathrm{dd}}^2f$~\cite{chandrakasan1992lowpower}. Reducing frequency lowers switching activity and can also permit lower voltage. Energy savings depend on execution time as well as power. Slowing one phase can keep both active and waiting GPUs drawing power for longer. Over a measurement interval of duration $T$, we define energy as $E=\int_0^T\sum_{g\in\mathcal{G}}P_g(t)\,dt$, throughput as $q=N/T$, and energy per valid training token as $e=E/N$, where $\mathcal{G}$ is the allocated GPU set, $P_g(t)$ is the power of GPU $g$, and $N$ counts valid training tokens. Energy includes every allocated GPU, whether active or waiting, but excludes CPU, network, and cooling energy.

\section{Characterizing Energy-Saving Headroom}
\label{sec:headroom}

We now quantify this headroom. We sweep reduced generation and training frequencies and compare their energy per valid training token and end-to-end throughput with the governor across workloads, GPU allocations, and execution modes.

\subsection{Where power can be reduced}
Figure~\ref{fig:governor-profile} compares the same workload and seed under the governor and a lower-frequency setting. Power falls in every phase, most during log-probability computation and training, while phase durations remain close. This example illustrates a setting where lower power outweighs the added execution time. The multi-seed sweeps below examine how consistently this trade-off holds across frequency pairs and workloads.

\subsection{How the opportunity varies across workloads}
We vary two frequency settings, referred to as generation and training frequencies. In colocated execution, these settings follow phase boundaries on shared GPUs. In disaggregated and asynchronous execution, each applies to its corresponding GPU pool. Each setting is compared with the governor using the same workload and GPU allocation size. Table~\ref{tab:workloads} summarizes these configurations.

\begin{table}[!t]
\caption{GRPO headroom workloads and GPU configurations for updates
5--25. Tuples give prompts per update $P$, responses per prompt $K$,
and training batch size $B$. GPU splits denote generation/training
allocations. Checkmarks indicate tested sizes and dashes indicate
sizes not included.}
\label{tab:workloads}
\label{tab:workloads-gb200}
\vspace{4pt}
\centering\footnotesize

\begin{tabular*}{\columnwidth}{
@{\extracolsep{\fill}}lccc@{}}
\toprule
Model & Small (S) & Medium (M) & Large (L)\\
\midrule
Qwen
& $(32,4,128)$ & $(64,8,512)$ & $(128,16,2048)$\\
Nemotron
& $(16,4,64)$ & $(16,8,128)$ & $(32,8,256)$\\
\bottomrule
\end{tabular*}
\par\smallskip
{\footnotesize Sequence limits: Qwen 4096; Nemotron 8192.\par}

\par\medskip
\textbf{(a) H100}\par\smallskip
\begin{tabular*}{\columnwidth}{
@{\extracolsep{\fill}}lcccc@{}}
\toprule
Deployment & GPUs & S & M & L\\
\midrule
\rowcolor[gray]{0.93}
\multicolumn{5}{c}{\textbf{Qwen3-30B-A3B}}\\
Colocated sync.
& 32, 64 & $\checkmark$ & $\checkmark$ & $\checkmark$\\
Disaggregated sync.
& 8/32, 16/32, 8/24
& $\checkmark$ & $\checkmark$ & $\checkmark$\\
Async, drained
& 16/32, 8/24 & $\checkmark$ & $\checkmark$ & ---\\
Async, in-flight
& 16/32, 8/24 & $\checkmark$ & $\checkmark$ & $\checkmark$\\

\addlinespace[3pt]
\rowcolor[gray]{0.93}
\multicolumn{5}{c}{\textbf{Nemotron-Super-49B}}\\
Colocated sync.
& 64 & $\checkmark$ & $\checkmark$ & $\checkmark$\\
Disaggregated sync.
& 16/32 & $\checkmark$ & $\checkmark$ & ---\\
& 16/64 & $\checkmark$ & $\checkmark$ & $\checkmark$\\
Async, drained
& 16/32, 16/64 & $\checkmark$ & $\checkmark$ & ---\\
Async, in-flight
& 16/32 & $\checkmark$ & $\checkmark$ & ---\\
& 16/64 & $\checkmark$ & $\checkmark$ & $\checkmark$\\
\bottomrule
\end{tabular*}

\par\medskip
\textbf{(b) GB200}\par\smallskip
\begin{tabular*}{\columnwidth}{
@{\extracolsep{\fill}}lcccc@{}}
\toprule
Deployment & GPUs & S & M & L\\
\midrule
\rowcolor[gray]{0.93}
\multicolumn{5}{c}{\textbf{Qwen3-30B-A3B}}\\
Colocated sync.
& 16 & $\checkmark$ & $\checkmark$ & $\checkmark$\\
Disaggregated sync.
& 8/16 & $\checkmark$ & $\checkmark$ & ---\\
& 12/24 & $\checkmark$ & $\checkmark$ & $\checkmark$\\
Async, drained
& 4/16 & --- & $\checkmark$ & ---\\
Async, in-flight
& 4/16 & --- & $\checkmark$ & ---\\

\addlinespace[3pt]
\rowcolor[gray]{0.93}
\multicolumn{5}{c}{\textbf{Nemotron-Super-49B}}\\
Colocated sync.
& 32 & $\checkmark$ & $\checkmark$ & $\checkmark$\\
\bottomrule
\end{tabular*}
\end{table}

Table~\ref{tab:headroom-summary} reports representative settings for every execution mode, with energy and throughput changes relative to the governor.

\begin{table*}[t]
\caption{Representative headroom relative to the governor, with
four matched seeds per row. Energy saving measures the reduction
in all-GPU energy per valid training token, and throughput measures
valid training tokens per second.}
\label{tab:headroom-summary}
\vspace{4pt}
\centering\footnotesize
\setlength{\tabcolsep}{5pt}
\renewcommand{\arraystretch}{1.10}

\textbf{(a) Synchronous execution}\par\smallskip
\begin{tabular*}{\textwidth}{
@{\extracolsep{\fill}}lccccc@{}}
\toprule
Model
& GPUs
& Workload
& \shortstack{Gen/train\\(MHz)}
& \shortstack{Energy saving\\(\%)}
& \shortstack{Throughput change\\(\%)}\\
\midrule
\rowcolor[gray]{0.93}
\multicolumn{6}{c}{\textbf{Colocated}}\\
Qwen     & 32 H100  & Large  & 1800/1500 & 8.29  & $-0.99$\\
Qwen     & 16 GB200 & Medium & 1875/1020 & 11.53 & $-0.69$\\
Nemotron & 32 GB200 & Medium & 1800/1500 & 5.08  & $-1.79$\\

\addlinespace[4pt]
\rowcolor[gray]{0.93}
\multicolumn{6}{c}{\textbf{Disaggregated}}\\
Qwen     & 8/32 H100   & Large  & 1500/1200 & 28.23 & $-0.36$\\
Nemotron & 16/64 H100  & Medium & 1350/1350 & 26.14 & $-1.16$\\
Nemotron & 16/64 H100  & Small  & 1350/1500 & 23.75 & $-1.31$\\
Qwen     & 12/24 GB200 & Small  & 1800/1400 & 14.31 & $-0.44$\\
Qwen     & 12/24 GB200 & Medium & 1800/1400 & 13.27 & $-0.47$\\
\bottomrule
\end{tabular*}

\par\medskip
\textbf{(b) Asynchronous execution}\par\smallskip
\begin{tabular*}{\textwidth}{
@{\extracolsep{\fill}}lcccccc@{}}
\toprule
Model
& GPUs
& Workload
& \shortstack{Gen/train\\(MHz)}
& \shortstack{Energy saving\\(\%)}
& \shortstack{Throughput change\\(\%)}
& \shortstack{$\Delta$ training\\reward}\\
\midrule
\rowcolor[gray]{0.93}
\multicolumn{7}{c}{
\textbf{Drained updates}\quad (policy age 1)}\\
Nemotron & 16/64 H100 & Medium & 1725/810  & 21.85 & $-1.83$ & $+0.0015$\\
Nemotron & 16/64 H100 & Medium & 1830/1200 & 21.13 & $-0.26$ & $-0.0038$\\
Qwen     & 8/24 H100  & Medium & 1830/810  & 19.89 & $-1.37$ & $-0.0006$\\
Qwen     & 4/16 GB200 & Medium & 1900/1700 & 7.84  & $-1.90$ & $+0.0005$\\

\addlinespace[4pt]
\rowcolor[gray]{0.93}
\multicolumn{7}{c}{
\textbf{In-flight updates}\quad (policy age 2)}\\
Nemotron & 16/64 H100 & Small  & 1725/1200 & 21.69 & $+1.48$ & $-0.0047$\\
Nemotron & 16/64 H100 & Medium & 1725/1200 & 21.27 & $-1.37$ & $+0.0052$\\
Qwen     & 8/24 H100  & Large  & 1830/810  & 19.91 & $-0.78$ & $-0.0009$\\
Qwen     & 4/16 GB200 & Medium & 1900/1400 & 8.91  & $-1.24$ & $+0.0015$\\
\bottomrule
\end{tabular*}

\par\smallskip
{\raggedright\footnotesize
Pool splits give generation/training GPU counts.
Each row includes its study's idle clock setting.
All rows cover updates 5--25.
$\Delta$ training reward is the signed difference in mean training
reward from the governor and serves as a learning-behavior diagnostic.
\par}
\end{table*}

Among the synchronous settings, colocated execution saves 5.08--11.53\% energy per valid training token across both platforms, with throughput reductions below 2\%. Disaggregated execution offers more headroom, because each pool waits while the other is busy and that waiting time can run at a lower clock. On H100, Qwen and Nemotron save 23.75--28.23\% with at most 1.31\% lower throughput. On GB200, disaggregated Qwen saves 14.31\% and 13.27\% on the small and medium workloads, with less than 1\% lower throughput.

\subsection{When generation overlaps training}
Asynchronous execution also makes the relative completion times of the generation and training pools important. A pool that repeatedly finishes before the next dependency may tolerate a lower frequency without reducing end-to-end throughput because its additional computation time can replace time otherwise spent waiting for the other pool.

The asynchronous rows of Table~\ref{tab:headroom-summary} cover drained updates with policy age 1 and in-flight updates with policy age 2. The selected drained and in-flight settings save up to 21.85\% and 21.69\% energy per valid training token, respectively, with measured throughput reductions below 2\%. Drained results also lower the clock of idle pools, which the governor does not do, so their savings combine active and idle clock control. For these rows, the table also reports training-reward differences as a learning-behavior diagnostic rather than evidence of held-out accuracy equivalence.

Across all modes, the most efficient pair depends on the workload and deployment, so a pair selected for one configuration should not be assumed suitable for another.
\section{Using Headroom to Reduce Energy}
\label{sec:opportunity}

The headroom measurements show that lower frequencies can save energy, but further reductions can eventually reverse the benefit. A useful frequency choice must balance lower power against longer execution, including the energy consumed by waiting GPUs.

\subsection{Balancing power reduction against slower progress}
When a phase is limited by memory movement or communication, additional core speed may do little to improve throughput. Reducing frequency can therefore lower power with little effect on completion time, consistent with the distinction between compute and bandwidth limits in the Roofline model~\cite{williams2009roofline}. At sufficiently low frequencies, however, slower computation can extend the phase and reduce end-to-end throughput.

The energy cost follows directly from the metrics in Section~\ref{sec:background}. With $\overline P$ denoting average power summed across the allocated GPUs,
\begin{equation}
 e=\frac{E}{N}=\frac{\overline P T}{N}=\frac{\overline P}{q}.
 \label{eq:power-rate}
\end{equation}
Energy per valid training token falls when power decreases proportionally more than throughput. A further frequency reduction becomes counterproductive when the additional execution time outweighs the power saving. The lowest available frequency is therefore not necessarily the most efficient choice.

\begin{figure*}[!t]
\centering\includegraphics[width=\textwidth]{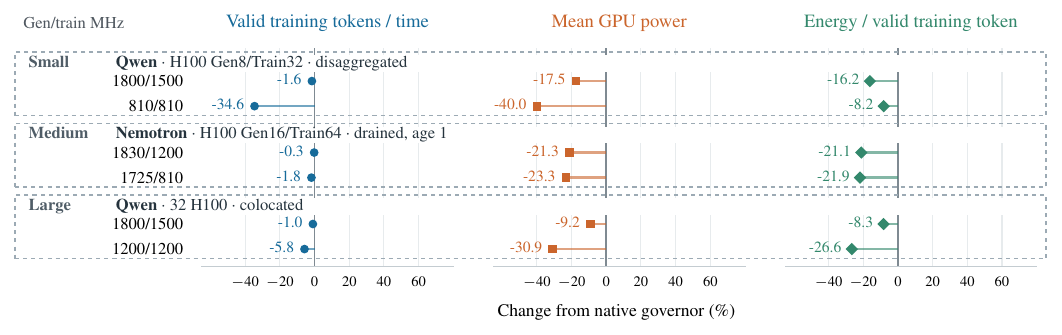}
\caption{Throughput, power, and energy-per-token changes relative to the governor for representative generation/training frequency pairs, as geometric means over four matched seeds.}
\label{fig:power-time-energy}
\end{figure*}

Figure~\ref{fig:power-time-energy} separates throughput loss, power reduction, and the resulting energy change. For disaggregated Qwen, the lowest-frequency setting reduces power further but slows execution enough to give back much of the saving. In contrast, the asynchronous Nemotron results show substantial energy savings with little throughput loss. These examples illustrate why frequency selection must consider both power and end-to-end throughput.

\subsection{Choosing a setting under a throughput limit}
We express the performance requirement as an allowed throughput loss relative to the governor. Among tested settings that meet this requirement and save energy, we select the one with the lowest measured energy per valid training token. A setting that saves more energy but exceeds the throughput-loss allowance is not eligible, and the governor remains available if no tested setting qualifies. Table~\ref{tab:control-settings} lists the throughput-loss allowance.

\section{EE-RL: Runtime GPU Frequency Control}
\label{sec:design}\label{sec:overview}
Choosing frequencies at runtime exploits energy-saving headroom without extensive profiling in advance. \system tests lower generation and training clocks in short trials against the governor
and selects the most energy-efficient setting that retains at least 98\% of governor throughput. It then monitors energy and throughput to guide adjustments as the workload changes.

Two questions shape the design. The first is how to judge a trial fairly when the work changes from update to update (Section~\ref{sec:design-work}), and the second is how to search when the stages interact differently in each execution mode (Sections~\ref{sec:design-select} and~\ref{sec:design-inflight}).

\subsection{Comparing at equal work}
\label{sec:design-work}
\begin{figure}[!t]
\centering
\includegraphics[width=\columnwidth]{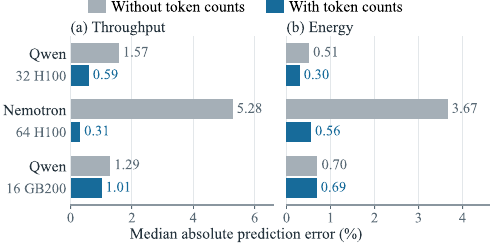}
\caption{Median absolute error in predicting governor throughput and energy for later updates, with and without token counts. Lower is better.}
\label{fig:prediction}
\end{figure}

An update with longer responses takes longer at any clock, so elapsed time alone would confuse a change in work with the effect of a frequency. \system instead asks what the governor would have done with the same work. Each run begins with a calibration phase, in which the first updates run at governor clocks and the controller only observes. From governor updates only, starting with these, it fits
\begin{equation}
\log \widehat y_A=\beta_0+\textstyle\sum_{k}\beta_k z_k,
\label{eq:token-prediction}
\end{equation}
where $\widehat y_A$ is a predicted governor time and each $z_k$ is the standardized $\log(1+x)$ of a token count $x$, which is the number of generated tokens $G$, valid training tokens $N$, or tokens in the longest response $L$. Generation time uses $G$ and $L$, since generation cannot finish before its longest response does. Training time uses $N$. When the stages run in sequence, update time is their sum plus remaining work such as refit and reward. When they overlap, it is the longer of the two. Figure~\ref{fig:prediction} tests this adjustment by predicting later governor updates from earlier ones. Token counts lower median prediction error in every measured configuration, most for H100 Nemotron. The gain is smallest for GB200 Qwen, particularly for energy.

\subsection{Choosing a frequency pair}
\label{sec:design-select}
In synchronous execution and with drained asynchronous updates, every update runs under a single clock setting, so a trial of a few updates can be judged directly against the governor prediction. The governor observations around each trial are added to the references, so the prediction draws on recent governor behavior. A candidate passes if it retains 98\% of predicted throughput and lowers energy per valid training token (Algorithm~\ref{alg:calibration}). Search follows the execution schedule: synchronous execution searches both phase clocks together, while drained execution searches the pools sequentially (Section~\ref{sec:implementation}).

\begin{algorithm}[!t]
\caption{Choosing GPU frequencies in synchronous and drained execution}
\label{alg:calibration}
\small
\begin{algorithmic}[1]
\REQUIRE Supported frequencies, governor observations
\STATE Fit the governor predictor on token counts and longest response
\FOR{each search round}
  \STATE \COMMENT{Synchronous: both clocks. Drained: first pool, then second pool.}
  \REPEAT
    \STATE Propose a candidate for this round
    \STATE Run it between governor observations
    \STATE Compare throughput and energy with the governor prediction
  \UNTIL{the round's search ends}
  \STATE Confirm the passing setting with the lowest energy per token on fresh observations
  \STATE If confirmation fails, try the next passing setting or continue the search
  \STATE Keep the confirmed setting fixed in later rounds
\ENDFOR
\STATE Lock the confirmed setting
\end{algorithmic}
\end{algorithm}

A setting that passes a short trial may become less suitable as the workload changes. In synchronous and drained execution, \system checks a trailing window of 20 usable updates every 10 updates, comparing throughput and energy with the governor reference for the same work. A throughput shortfall prompts a clock increase, while sufficient margin permits a further reduction. Each clock change starts a fresh lookback window. Workload drift or repeated coverage failures trigger new governor measurements to refresh the reference.

When synchronous or drained execution places generation and training on separate GPU pools, one pool often waits for the other before its next phase can start. The controller drops the waiting pool to an idle clock until that phase begins. No computation runs in these gaps, so this saves energy that the governor spends on idle GPUs, without slowing the update.

\subsection{Choosing frequencies with in-flight updates}
\label{sec:design-inflight}
With in-flight updates, generation continues across updates, pausing only for each weight refit. Requests started under one clock can finish under the next, so the in-flight controller, unlike the drained one, cannot fit generation time at a given clock. \system therefore treats selection as an experiment. Each trial places a block of candidate updates between governor blocks and compares their throughput and energy at matched work. Block length is set from the noise measured during calibration. As with drained updates, pools are searched one at a time, beginning with the less busy one, and the best setting is confirmed before it is locked.

Because a single confirmation cannot rule out a small loss, the controller keeps one governor block for every two candidate blocks after locking, so the comparison with the governor continues.

\section{Control Across Execution Modes}
\label{sec:implementation}
A slower phase does not always mean a slower RL update. In synchronous execution, added computation time extends the update. With drained updates, it may replace time spent waiting for the other pool. In-flight responses can cross a clock change, making their cost hard to attribute to either setting. These dependencies determine how \system measures trials and searches frequencies. Table~\ref{tab:control-settings} gives the configuration.

\begin{table}[!t]
\caption{Controller configuration.}
\label{tab:control-settings}
\vspace{3pt}
\centering\footnotesize
\setlength{\tabcolsep}{3pt}
\renewcommand{\arraystretch}{1.15}

\begin{tabularx}{\columnwidth}{
@{}>{\raggedright\arraybackslash}p{.23\columnwidth}
*{3}{>{\centering\arraybackslash}X}@{}}
\toprule
Setting & Sync. & Drained & In-flight\\
\midrule
Allowed loss
& 2\% & 2\% & 2\%\\
Calibration
& 15 updates & 15 updates & 45 updates\\
Trial
& 3 updates & 3 updates & 3, 6, or 12 updates\\
Confirmation
& 6 fresh updates & 6 fresh updates & 1--2 trials\\

\addlinespace[3pt]
Search grid
& 90\,MHz & 50/100\,MHz & 50/100\,MHz\\
Feedback step
& 45/90\,MHz & 50\,MHz & ---\\
Monitoring
& \shortstack{20 updates,\\every 10}
& \shortstack{20 updates,\\every 10}
& 1 governor per 2 candidate blocks\\
Idle clock
& 345\,MHz & 345\,MHz & ---\\
\bottomrule
\end{tabularx}

\par\smallskip
{\raggedright\footnotesize
Allowed loss is relative to governor throughput.
Asynchronous search uses 50\,MHz spacing for the first pool
and 100\,MHz for the second.
Synchronous feedback uses 45\,MHz steps, increasing to 90\,MHz
above 4\% throughput loss. 
GPUs support only discrete clock frequencies, so applied frequency steps can differ slightly from these nominal sizes.
The idle clock applies only to separate GPU pools.
\par}
\end{table}

\subsection{Synchronous execution}
\label{sec:runtime-sync}
Generation and training run in sequence, so their added execution times share the same throughput allowance. \system fits a governor predictor for each phase and adds the median remaining time, including refit and reward. During a trial, it measures each phase's time and energy relative to its governor prediction at the observed token workload. Applying these ratios to the governor reference updates estimates the candidate's throughput and energy over a common workload, reducing sensitivity to the few updates in a trial.

The phase timings identify which clock needs to rise when a pair is too slow. Search starts with both clocks reduced and raises the phase responsible for most of the added time. Once a pair passes, it explores lower settings for further energy savings.

\subsection{Asynchronous execution with drained updates}
\label{sec:runtime-drain}
Generation and training overlap but must rendezvous before each weight update, and clocks change at this rendezvous. The later arrival determines update time, so slower computation in the earlier pool may simply shorten its wait. \system records both arrivals and searches the pool that usually arrives first, leaving the other on the governor. Its confirmed setting then provides the reference for lowering the second pool. The combined pair must meet the throughput requirement and use no more energy than the first setting alone, so a second lowered clock is kept only if it benefits the whole allocation.

Response-length variation can obscure this waiting interval. When only training is slowed, \system applies the measured slowdown to the governor references and recomputes update time from their recorded arrivals. Holding generation fixed estimates how much training can slow before delaying the rendezvous. Other candidates use the whole-update governor prediction.

\subsection{Asynchronous execution with in-flight updates}
\label{sec:runtime-inflight}
A response may begin under one clock and finish under another. Measuring the responses completed within an update would therefore mix work from different settings. The runtime instead counts the valid tokens consumed by training, changes clocks during the weight-refit pause, and leaves updates carrying requests from the preceding setting out of trial decisions. The longest consumed response remains a predictor feature because it improves held-out estimates when response lengths vary, even though it no longer determines an update boundary.

Neighboring updates also share work, so a short trial may contain less independent evidence than its length suggests. Calibration estimates this correlation and selects the shortest trial block meeting a 2\% standard-error target for comparison with the governor. Each candidate block is bracketed by governor blocks. Without a drained rendezvous to measure waiting directly, search starts with the less busy pool, then tests the second with the first setting held. Governor blocks continue after locking to check the throughput requirement.

\subsection{Runtime measurement}
\label{sec:telemetry}
The energy accounting must include waiting GPUs: slowing one pool can extend the time the other remains powered. The runtime samples each GPU every 200\,ms and sums interpolated NVML energy counters across the allocation~\cite{nvidiaNvmlUtilization}. Acceptance uses whole-update time and energy, including waits and clock changes.

\begin{table*}[t]
\caption{Controller performance relative to matched governor runs over the listed updates, including search and adaptation costs. Energy savings and throughput changes use valid training tokens. Accuracy change is the difference in final validation accuracy from the governor, in percentage points. Dashes indicate unreported accuracy. GPU splits give generation/training counts. Drained and in-flight runs use policy ages 1 and 2, respectively.}
\label{tab:controller}
\centering\footnotesize
\setlength{\tabcolsep}{3pt}
\renewcommand{\arraystretch}{1.08}

\begin{tabular*}{\textwidth}{@{\extracolsep{\fill}}llcrrrrr@{}}
\toprule
Model / execution
& GPUs
& $(P,K,B)$
& \shortstack{Token\\limit}
& Updates
& \shortstack{Energy/token\\saving (\%)}
& \shortstack{Throughput\\change (\%)}
& \shortstack{Accuracy\\change (pp)}\\
\midrule
Qwen, colocated sync.
& 32 H100 & 64/8/512 & 4096 & 2--107
& 7.00 & $-1.25$ & ---\\
Qwen, disaggregated sync.
& 8/32 H100 & 64/8/512 & 4096 & 2--93
& 16.94 & $-1.39$ & ---\\
Mistral, colocated sync.
& 16 H100 & 64/8/512 & 4096 & 2--265
& 4.14 & $+1.62$ & ---\\
Mistral, disaggregated sync.
& 8/16 H100 & 64/8/512 & 4096 & 2--315
& 8.89 & $-1.52$ & ---\\
Qwen, async drained
& 8/24 H100 & 32/4/128 & 4096 & 2--175
& 14.51 & $-1.97$ & $0.00$\\
Mistral, async drained
& 8/16 H100 & 64/8/512 & 4096 & 2--300
& 13.56 & $+1.46$ & $+0.90$\\
Qwen, async in-flight
& 16/32 H100 & 32/4/128 & 4096 & 2--335
& 7.24 & $-1.08$ & $+0.18$\\
Mistral, async in-flight
& 8/16 H100 & 64/8/512 & 4096 & 2--560
& 6.52 & $-1.61$ & $-0.14$\\
\bottomrule
\end{tabular*}
\end{table*}

\section{Evaluation}
\label{sec:evaluation}
\system saves energy in all eight controller comparisons after including selection and adaptation costs. We examine how the execution schedule determines the useful control target and why the savings after locking differ from those over the complete interval.

\subsection{Experimental setup}
We evaluate Qwen3-30B-A3B and Mistral-NeMo 12B on H100 in colocated and disaggregated synchronous execution, and with drained or in-flight asynchronous updates. Mistral-NeMo was not part of the headroom study, so it tests the controller on a model without a prior sweep. Nemotron-Super-49B's longer updates left too few steps within the allocated runtime for controller evaluation.

Each comparison matches the seed, workload, and GPU topology, using separate physical allocations. Rates use summed valid training tokens, time, and all-GPU energy over the same updates, including search, confirmation, correction, waiting, and synchronization. Table~\ref{tab:controller} reports every configuration. RL workloads also vary from run to run, even with matched seeds, because sampling and the policy drift apart and response lengths change. We therefore also compare each controller run with a governor model fitted on two pooled governor runs of the same workload, and report these workload-normalized results in Appendix~\ref{supp:normalized}.

\subsection{Savings survive online selection}
Disaggregated synchronous Qwen achieves the largest saving: 16.9\% with 1.4\% lower throughput, compared with 7.0\% energy saving on colocated. Mistral-NeMo shows the same ordering. Separate pools expose idle intervals in addition to headroom during computation, and the controller reduces clocks in both pools.

The Mistral-NeMo results show that useful settings can be found without a prior headroom sweep. Their savings still depend on the balance between the pools and their execution schedule.

\subsection{Overlap makes training the useful control target}
In all four asynchronous runs, the controller tests lower generation clocks but ultimately slows only training. These choices suggest that training can run at a lower frequency while generation prepares the next batch, whereas slowing generation does not yield an accepted energy--throughput trade-off. Arrival order or utilization determines which GPU pool is searched first, but the controller's decisions depend on how each setting affects end-to-end progress.

In drained execution, this spare training capacity translates into energy savings of about 14\% over the full interval and about 20\% after locking. Slower training uses time that would otherwise be spent waiting at the rendezvous. Lowering clocks during idle periods adds to the savings for Qwen, but rarely triggers for Mistral-NeMo because its idle gaps are short. The two workloads thus achieve similar savings through different combinations of active and idle frequency control, with the benefit measured across the whole GPU allocation.

\begin{table*}[t]
\caption{Controller performance before and after locking, relative to the paired governor over the same updates. Energy savings are measured per valid training token. All runs use H100 GPUs. GPU splits give generation/training counts. Gov denotes native governor control, and Varies denotes changing frequency settings. After-lock intervals include subsequent adaptation and, for in-flight runs, governor comparison blocks.}
\label{tab:controller-progress}
\centering\footnotesize
\setlength{\tabcolsep}{3pt}
\renewcommand{\arraystretch}{1.0}

\begin{tabular*}{\textwidth}{@{\extracolsep{\fill}}llrcrr@{}}
\toprule
Model / execution
& Interval
& Updates
& \shortstack{Gen/train\\(MHz)}
& \shortstack{Energy/token\\saving (\%)}
& \shortstack{Throughput\\change (\%)}\\
\midrule

Qwen, colocated sync.
& Before lock & 2--77 & Varies & 6.59 & $-1.13$\\
& After lock & 78--107 & 1830/1425 & 8.04 & $-1.55$\\
\hdashline[0.5pt/2pt]

Qwen, disaggregated sync.
& Before lock & 2--61 & Varies & 14.02 & $-1.27$\\
& After lock & 62--93 & 1695/1425 & 22.44 & $-1.60$\\
\hdashline[0.5pt/2pt]

Mistral, colocated sync.
& Before lock & 2--172 & Varies & 3.95 & $-0.18$\\
& After lock & 173--265 & 1785/1695 & 4.56 & $+4.99$\\
\hdashline[0.5pt/2pt]

Mistral, disaggregated sync.
& Before lock & 2--59 & Varies & 8.04 & $-1.01$\\
& After lock & 60--315 & Varies & 9.08 & $-1.64$\\
\hdashline[0.5pt/2pt]

Qwen, async drained, 8/24
& Before lock & 2--89 & Varies & 9.91 & $-2.96$\\
& After lock & 90--175 & Gov/Varies & 19.23 & $-0.94$\\
\hdashline[0.5pt/2pt]

Mistral, async drained, 8/16
& Before lock & 2--190 & Varies & 9.32 & $-0.10$\\
& After lock & 191--300 & Gov/1185 & 20.87 & $+4.19$\\
\hdashline[0.5pt/2pt]

Qwen, async in-flight, 16/32
& Before lock & 2--257 & Varies & 6.45 & $-0.89$\\
& After lock & 258--335 & Gov/1125 & 9.82 & $-1.70$\\
\hdashline[0.5pt/2pt]

Mistral, async in-flight, 8/16
& Before lock & 2--313 & Varies & 5.32 & $-2.45$\\
& After lock & 314--560 & Gov/1425 & 8.03 & $-0.52$\\

\bottomrule
\end{tabular*}
\end{table*}

\subsection{Selection and monitoring have different costs}
\begin{figure}[!t]
\centering\includegraphics[width=\columnwidth]{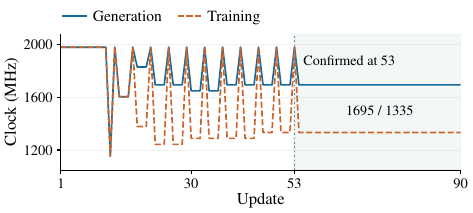}
\caption{Measured phase clocks in a separate 90-update colocated Qwen run on 32 H100s. Confirmation after update 53 selects different generation and training frequencies. This trace illustrates search and is separate from the scored runs.}
\label{fig:controller-trace}
\end{figure}

Every run saves more energy after locking than before it  (Table~\ref{tab:controller-progress}). Disaggregated synchronous Qwen saves 22.4\% after locking, versus 16.9\% over the complete interval. Figure~\ref{fig:controller-trace} shows this selection process in a separate colocated run, where candidate trials alternate with governor observations before confirmation. Reporting only the after-lock interval would miss these selection costs.

After locking, monitoring both recovers throughput and finds further reductions. For disaggregated synchronous Mistral-NeMo, three clock increases are followed by lower throughput losses: 3.30\% to 1.65\%, 2.35\% to 1.20\%, and 2.70\% to 0.34\% in the next 20-update windows. These windows retain 9.06--9.66\% energy savings in the controller's work-normalized estimates. For drained Qwen, monitoring identifies sufficient throughput margin to lower the training clock from 1065 to 1005\,MHz. Following confirmation, this setting achieves 19.50\% energy savings with 0.56\% throughput loss against the paired governor over updates 142--175.

In-flight execution also pays for measurement after selection. Its full-interval savings are roughly half those of drained execution. Noisier measurements delay locking, and about a third of after-lock updates still use the governor for continued comparison. In-flight runs also omit idle dropping. The different schedules and allocations prevent attributing the gap to any one factor, but the persistent governor blocks explain why a longer run alone would not recover all of the available saving. Longer stable runs can amortize a completed search, but not a fixed fraction of time spent rechecking the choice.

\subsection{Interpreting throughput and learning quality}
All eight full-interval comparisons lose less than 2\% throughput, although two asynchronous runs exceed that loss before locking  (Table~\ref{tab:controller-progress}). A locally accepted setting therefore does not guarantee a run-level bound that includes unsuccessful trials and corrections.

Workload differences also matter. Synchronous Mistral-NeMo controller runs consume more valid tokens, and drained Mistral-NeMo has shorter longest responses after locking. These changes contribute to throughput gains and energy-per-token savings, so they should not be read as clock-induced speedups. The asynchronous validation-accuracy changes in Table~\ref{tab:controller} are all below one percentage point in magnitude, but one matched seed per configuration cannot establish equal learning quality.

\section{Discussion and Limitations}
\textbf{Controlling the final throughput cost.} The difference between trial-level acceptance and complete-run throughput suggests managing performance as an accumulated budget. A controller could track estimated accumulated slowdown and use the remaining allowance to limit the duration of new trials. This would connect exploration cost to the job's remaining horizon. Early in a run, more exploration may be economical. For production runs spanning thousands of updates, the longer horizon could justify a more extensive search: its upfront cost can be amortized over many subsequent updates, potentially allowing the controller to find more energy-efficient settings and achieve larger net savings. Near completion, retaining an accepted setting may contribute more than another search step.

\textbf{Workload and allocation dependence.} A frequency choice depends on token counts, response lengths, the work within each stage, and the GPU allocation. The stages themselves depend on the RL algorithm. We evaluate only GRPO, while methods that train a value model, such as PPO~\cite{schulman2017ppo}, add stages with their own frequency demands. Changes in pool size or parallelism can shift the balance between stages, and we also observed governor-performance differences across physical allocations. Resuming a checkpoint on a different allocation should therefore trigger a fresh frequency search. Our controller evaluation uses one matched seed per configuration. Repeated seeds and component ablations would quantify variability and separate the contributions of prediction, search, confirmation, and feedback. Replaying a fixed pair over the same horizon would directly compare adaptation with static control.

\textbf{Learning quality.} Across the four asynchronous controller runs, final validation accuracy differs from the paired governor by between $-$0.14 and +0.90 percentage points. These differences are small, but each comparison uses one matched seed. Asynchronous timing can change which responses training consumes and their policy ages. Repeated runs across seeds would better characterize the effect on learning quality, and comparisons at a common held-out quality target would test whether the energy savings persist when matching achieved quality.

\textbf{Pool allocation and idle power.}
The benefit of frequency control depends on the balance between GPU pools. Slowing one pool can reduce its active power while extending the other pool's wait. Idle control can reduce the energy consumed during that wait, but its additional benefit depends on the pool's existing idle power. Whole-allocation energy captures both effects. Results that combine active-frequency control with idle dropping therefore measure both mechanisms together. They do not isolate the contribution of either mechanism.

\section{Related Work}
\label{sec:related}

The balance between power and progress also arises in conventional model training and in serving, where prior systems adjust GPU power, frequency, or scheduling to improve efficiency. These efforts complement work on RL systems that improves resource use through placement, parallelism, and scheduling. Together, they provide the context for \system's approach to frequency control during RL post-training.

\textbf{Energy-aware training.} Zeus jointly tunes batch size and GPU power limit for recurring training jobs through online exploration~\cite{you2023zeus}. EnvPipe and Perseus lower the frequency of computation off the critical path of pipeline-parallel training~\cite{choi2023envpipe,chung2024perseus}, and kernel-level DVFS selects clock frequencies for individual kernels~\cite{spaan2026kernel}. Zeus uses short iterative profiles to estimate power and throughput, while Perseus constructs frequency schedules over a static per-iteration computation graph. In RL post-training, generation and training respond differently to frequency. Response-length changes can alter execution time independently of frequency, and slowing one stage can extend the other stage's waiting time. \system therefore accounts for response-length changes when comparing settings, measures the energy of the whole allocation, including waiting GPUs, and includes exploration costs in its savings.

\textbf{Energy-aware LLM serving.} DynamoLLM reconfigures instance count, parallelism, and GPU frequency for pools of similar requests under latency targets~\cite{stojkovic2025dynamollm}, and throttLL'eM sets frequency from projected KV-cache usage and batch size~\cite{throttllem2024}. With prefill and decode disaggregated onto separate GPU pools~\cite{zhong2024distserve}, GreenLLM, VoltanaLLM, and DualScale set a separate frequency for each phase~\cite{greenllm2025,yu2025voltanallm,basit2026dualscale}, and AFlex further separates attention and feed-forward frequencies~\cite{hu2026aflex}. These systems optimize inference energy under serving latency constraints. \system instead controls the coupled generation and training stages of RL post-training, targeting whole-allocation energy per valid training token under an end-to-end throughput allowance. Serving-engine optimizations such as iteration-level scheduling~\cite{yu2022orca} and paged KV-cache management~\cite{kwon2023pagedattention} are complementary. They improve scheduling and memory efficiency and can change the frequency settings that are beneficial.

\textbf{RL post-training systems.} HybridFlow supports flexible placement and execution of RL dataflows~\cite{sheng2024hybridflow}, ReaL reallocates parameters to adapt parallelization across the stages of an RL step~\cite{mei2025real}, and HetRL schedules RL across heterogeneous GPUs~\cite{he2026hetrl}. RollPacker mitigates long-tail rollouts in synchronous RL post-training~\cite{rollpacker2025}. AReaL decouples generation from training while controlling data staleness~\cite{fu2025areal}, and PipelineRL and Laminar also run the two concurrently~\cite{piche2025pipelinerl,laminar2025}. These systems primarily improve throughput or learning speed through placement, parallelism, and scheduling rather than GPU frequency control. Measurements of Olmo 3 32B Think attribute most of its Think-specific post-training energy to rollout generation~\cite{morrison2026hidden}. \system instead targets energy by controlling clocks within a given placement, using phase boundaries and pool balance to identify headroom. Changes to placement or parallelism can alter phase balance and therefore warrant re-evaluating the selected frequencies.
\section{Conclusion}
Generation and training expose different opportunities for GPU frequency reduction during RL post-training. Measurements of GRPO workloads on H100 and GB200 show that these opportunities can reduce GPU energy per valid training token while largely preserving governor throughput. \system selects frequencies during execution, accounts for response-length variation, and revisits its choices as the workload changes. Phase-aware frequency control complements RL scheduling and placement optimizations by adapting GPU speed to the work being performed.

\section*{Acknowledgments}
We thank Trivikram Krishnamurthy, Sridutt Bhalachandra, Terry Kong, and Bhargava Gopireddy for their support of this work.

\bibliography{references}
\bibliographystyle{mlsys2025}
\section*{AI Usage Disclosure}
AI-based tools assisted with manuscript writing and editing, experimental and analysis scripts, and result consistency checks. The authors take responsibility for the final content and findings.
\clearpage
\onecolumn
\appendix
\raggedbottom
\clubpenalty=10000
\widowpenalty=10000
\setlength{\LTpre}{3pt}
\setlength{\LTpost}{3pt}
\setcounter{figure}{0}
\setcounter{table}{0}
\setcounter{equation}{0}
\renewcommand{\thefigure}{A\arabic{figure}}
\renewcommand{\thetable}{A\arabic{table}}
\renewcommand{\theequation}{A\arabic{equation}}
\renewcommand{\theHfigure}{appendix.\arabic{figure}}
\renewcommand{\theHtable}{appendix.\arabic{table}}
\renewcommand{\theHequation}{appendix.\arabic{equation}}

\section*{Appendix}
This appendix expands the headroom, token-work prediction, and controller experiments with measurement details, the complete valid headroom inventory, and workload-normalized controller results. 

\section{Estimation and Selection}
\label{supp:measurement-protocol}
For seed $s$ and setting $p$, form $q_{s,p}=\sum N/\sum T$ and $e_{s,p}=\sum E/\sum N$, where $N$ counts valid training tokens. For each matched seed let $x_s=\log(q_{s,p}/q_{s,A})$. The tables report
\[
\widehat R_q=\exp(\bar x).
\]
The energy ratio uses the same construction. Energy saving is $100(1-R_e)$, and throughput change is $100(R_q-1)$. Selection targets lower energy per token while retaining at least 98\% of governor throughput. 

\section{Accounting for Token Work and Response Length}

\begin{figure}[H]
\centering
\includegraphics[width=\textwidth]{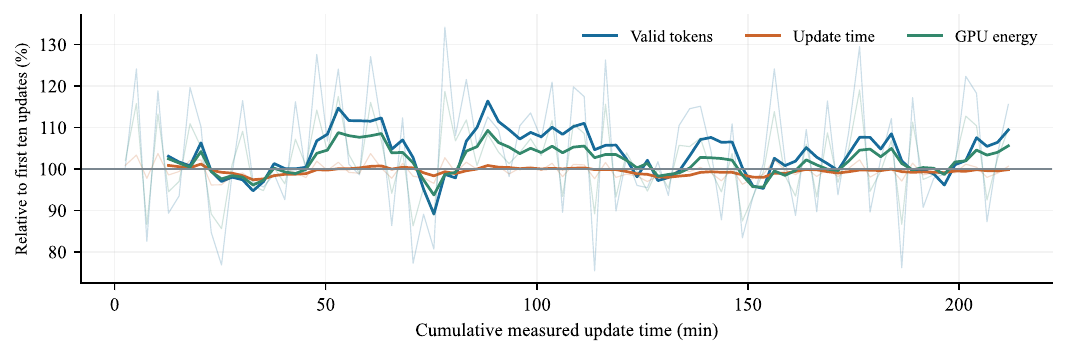}
\caption{Workload variation under governor-only Nemotron-Super-49B execution on 64 H100s. Faint lines show individual updates, and solid lines show trailing five-update means. Values are normalized to their mean over the first ten measured updates. Elapsed time sums measured update durations.}
\label{fig:governor-token-drift}
\end{figure}

A fixed number of RL updates does not imply a fixed amount of work. Longer responses increase the tokens processed by generation and training, while a particularly long response can delay generation even when total token counts are similar. Figure~\ref{fig:governor-token-drift} shows that work, time, and energy vary even under unchanged governor control. Without accounting for this variation, a frequency trial can appear faster or slower simply because its workload differs.

\system therefore predicts governor performance for the work observed during each candidate trial. The synchronous and drained references both use log-transformed generated and valid training token counts, together with the maximum response length, which governs when generation finishes. The in-flight reference uses the valid training tokens and longest response of the batch that training consumes. These models fit only governor observations, keeping candidate measurements from changing their own reference.

To evaluate token-count normalization, we use nine governor-only colocated runs: three seeds each for Qwen on 32 H100s, Nemotron on 64 H100s, and Qwen on 16 GB200s. Each forward test fits 15 past observations, predicts the next three, and advances by three, producing 88, 69, and 80 blocks, respectively. Compared with a constant reference, normalization reduces median absolute throughput prediction error from 1.57\% to 0.59\%, 5.28\% to 0.31\%, and 1.29\% to 1.01\%. Update-energy prediction errors also decrease.
\begin{center}\small\setlength{\tabcolsep}{5pt}
\begin{tabular}{lrrrrr}\toprule
Configuration & Blocks & \multicolumn{2}{c}{Throughput error (\%)} & \multicolumn{2}{c}{Energy error (\%)}\\
 & & Constant & Token-work & Constant & Token-work\\\midrule
Qwen, 32 H100 & 88 & 1.57 & 0.59 & 0.51 & 0.30\\
Nemotron, 64 H100 & 69 & 5.28 & 0.31 & 3.67 & 0.56\\
Qwen, 16 GB200 & 80 & 1.29 & 1.01 & 0.70 & 0.69\\
\bottomrule\end{tabular}\end{center}

A separate governor-only Mistral-NeMo 12B experiment tests the contribution of response length. Across 24 held-out three-update blocks, adding a longest-response proxy to the token-count model reduces update-time prediction RMS log error from 0.085 to 0.044. Work normalization improves the reference but does not eliminate the need for repeated measurements and continued monitoring.

\Needspace{5\baselineskip}
\section{Complete Synchronous Headroom Results}
\noindent\textbf{Note on GB200 measurements.} We observed substantial run-to-run variation on the GB200 cluster. The GB200 results in the synchronous and asynchronous tables below, as well as the GB200 rows of Table~\ref{tab:headroom-summary}, should be read with this variation in mind.

The tables contain 208 H100 and 85 GB200 non-governor comparisons, grouped by model and execution mode with hardware identified in each heading. Each uses a matched native-governor reference. Clock pairs are requested frequencies. Observed clocks may be slightly lower, and ``Gov'' leaves the phase governed. All synchronous comparisons aggregate four matched seeds. Row suffixes specify measured updates when they differ from updates 5--25.
\Needspace{10\baselineskip}
\subsection{Colocated synchronous: Qwen, 32 H100}
\noindent Four matched seeds per comparison.
{\small\setlength{\tabcolsep}{4pt}
\begin{longtable}{llrrl}
\toprule
Workload / split & Gen/train MHz & $\Delta q$ (\%) & Energy saving (\%) & $n$\\\midrule\endhead
Small & 1200/1200 & -12.10 & 20.91 & 4\\
Small & 1500/1500 & -4.84 & 9.56 & 4\\
Small & 1800/1800 & -0.80 & -0.13 & 4\\
Small & 1980/1980 & +0.39 & -0.70 & 4\\
Small & 1650/1200 & -4.85 & 11.85 & 4\\
Small & 1650/1350 & -3.24 & 10.03 & 4\\
Small & 1800/1200 & -2.27 & 9.97 & 4\\
Small & 1800/1350 & -2.54 & 7.45 & 4\\
Small & 1800/1500 & -1.68 & 4.69 & 4\\
Medium & 1200/1200 & -10.17 & 24.05 & 4\\
Medium & 1500/1500 & -4.67 & 10.88 & 4\\
Medium & 1800/1800 & -1.50 & 0.15 & 4\\
Medium & 1980/1980 & -1.21 & -1.23 & 4\\
Medium & 1650/1200 & -5.43 & 14.42 & 4\\
Medium & 1650/1350 & -3.89 & 12.48 & 4\\
Medium & 1800/1200 & -4.28 & 12.00 & 4\\
Medium & 1800/1350 & -3.54 & 9.44 & 4\\
Medium & 1800/1500 & -2.28 & 5.50 & 4\\
Large & 1200/1200 & -5.82 & 26.59 & 4\\
Large & 1500/1500 & -1.95 & 12.74 & 4\\
Large & 1800/1800 & -0.14 & 0.25 & 4\\
Large & 1980/1980 & -0.10 & -0.34 & 4\\
Large & 1650/1200 & -3.17 & 18.88 & 4\\
Large & 1650/1350 & -2.31 & 15.60 & 4\\
Large & 1800/1200 & -2.99 & 16.57 & 4\\
Large & 1800/1350 & -2.03 & 13.51 & 4\\
Large & 1800/1500 & -0.99 & 8.29 & 4\\
\bottomrule\end{longtable}}
\Needspace{10\baselineskip}
\subsection{Colocated synchronous: Qwen, 64 H100}
\noindent Four matched seeds per comparison.
{\small\setlength{\tabcolsep}{4pt}
\begin{longtable}{llrrl}
\toprule
Workload / split & Gen/train MHz & $\Delta q$ (\%) & Energy saving (\%) & $n$\\\midrule\endhead
Small & 1200/1200 & -14.16 & 18.26 & 4\\
Small & 1500/1500 & -5.57 & 8.52 & 4\\
Small & 1800/1800 & -1.84 & -0.51 & 4\\
Small & 1980/1980 & -1.84 & -1.71 & 4\\
Small & 1650/1200 & -6.59 & 10.05 & 4\\
Small & 1650/1350 & -5.29 & 8.33 & 4\\
Small & 1800/1200 & -4.76 & 7.63 & 4\\
Small & 1800/1350 & -3.83 & 6.20 & 4\\
Small & 1800/1500 & -2.93 & 3.35 & 4\\
Medium & 1200/1200 & -11.44 & 22.15 & 4\\
Medium & 1500/1500 & -4.53 & 10.14 & 4\\
Medium & 1800/1800 & -1.39 & 0.06 & 4\\
Medium & 1980/1980 & -1.50 & -1.37 & 4\\
Medium & 1650/1200 & -5.24 & 12.67 & 4\\
Medium & 1650/1350 & -4.15 & 10.75 & 4\\
Medium & 1800/1200 & -4.47 & 9.96 & 4\\
Medium & 1800/1350 & -2.86 & 8.34 & 4\\
Medium & 1800/1500 & -2.34 & 4.37 & 4\\
Large & 1200/1200 & -7.10 & 25.70 & 4\\
Large & 1500/1500 & -2.74 & 12.15 & 4\\
Large & 1800/1800 & -0.36 & 0.62 & 4\\
Large & 1980/1980 & -0.48 & -0.73 & 4\\
Large & 1650/1200 & -3.45 & 17.35 & 4\\
Large & 1650/1350 & -2.50 & 14.31 & 4\\
Large & 1800/1200 & -3.02 & 14.76 & 4\\
Large & 1800/1350 & -2.16 & 11.71 & 4\\
Large & 1800/1500 & -1.43 & 7.13 & 4\\
\bottomrule\end{longtable}}
\Needspace{10\baselineskip}
\subsection{Colocated synchronous: Qwen, 16 GB200}
\noindent Four matched seeds per comparison.
{\small\setlength{\tabcolsep}{4pt}
\begin{longtable}{llrrl}
\toprule
Workload / split & Gen/train MHz & $\Delta q$ (\%) & Energy saving (\%) & $n$\\\midrule\endhead
Small & 1875/1020 & +1.62 & 12.61 & 4\\
Small & 1875/930 & +1.53 & 11.98 & 4\\
Small & 1950/1020 & +4.35 & 12.62 & 4\\
Small & 1950/930 & +3.43 & 12.02 & 4\\
Small & Gov/1020 & +2.51 & 9.88 & 4\\
Small & Gov/840 & +2.02 & 10.40 & 4\\
Small & Gov/885 & +4.67 & 12.12 & 4\\
Small & Gov/930 & -0.98 & 9.12 & 4\\
Small & Gov/975 & -3.49 & 6.95 & 4\\
Medium & 1875/1020 & -0.69 & 11.53 & 4\\
Medium & 1875/930 & +1.05 & 12.82 & 4\\
Medium & 1950/1020 & -2.19 & 10.36 & 4\\
Medium & 1950/930 & +3.10 & 14.55 & 4\\
Medium & Gov/1020 & +11.44 & 16.99 & 4\\
Medium & Gov/840 & +2.38 & 13.71 & 4\\
Medium & Gov/885 & +0.03 & 10.41 & 4\\
Medium & Gov/930 & +6.19 & 14.64 & 4\\
Medium & Gov/975 & +0.80 & 11.90 & 4\\
Large & 1875/1020 & -1.08 & 13.44 & 4\\
Large & 1875/930 & +0.46 & 14.46 & 4\\
Large & 1950/1020 & +0.63 & 13.62 & 4\\
Large & 1950/930 & -7.65 & 8.78 & 4\\
Large & Gov/1020 & -3.30 & 10.03 & 4\\
Large & Gov/840 & -3.95 & 12.19 & 4\\
Large & Gov/885 & -5.42 & 12.05 & 4\\
Large & Gov/930 & -9.00 & 6.65 & 4\\
Large & Gov/975 & +1.35 & 13.65 & 4\\
\bottomrule\end{longtable}}
\Needspace{10\baselineskip}
\subsection{Colocated synchronous: Nemotron, 64 H100}
\noindent Four matched seeds per comparison.
{\small\setlength{\tabcolsep}{4pt}
\begin{longtable}{llrrl}
\toprule
Workload / split & Gen/train MHz & $\Delta q$ (\%) & Energy saving (\%) & $n$\\\midrule\endhead
Small & 1200/1200 & -11.87 & 21.33 & 4\\
Small & 1500/1500 & -3.26 & 8.32 & 4\\
Small & 1800/1800 & -0.41 & 0.03 & 4\\
Small & 1980/1980 & -1.80 & -1.41 & 4\\
Small & 1650/1200 & -4.38 & 7.51 & 4\\
Small & 1650/1350 & -3.64 & 6.40 & 4\\
Small & 1800/1200 & -3.64 & 4.54 & 4\\
Small & 1800/1350 & -2.03 & 3.75 & 4\\
Small & 1800/1500 & -1.12 & 1.46 & 4\\
Small & Gov/1500 & -0.73 & 1.88 & 4\\
Small & Gov/1350 & -2.52 & 3.19 & 4\\
Small & 1800/Gov & -0.80 & 0.04 & 4\\
Small & 1650/Gov & -2.98 & 3.03 & 4\\
Small & Gov/1200 & -2.24 & 4.77 & 4\\
Small & 1800/1650 & -0.63 & 0.78 & 4\\
Medium & 1200/1200 & -10.92 & 22.48 & 4\\
Medium & 1500/1500 & -3.82 & 8.97 & 4\\
Medium & 1800/1800 & -0.75 & -0.10 & 4\\
Medium & 1980/1980 & -0.77 & -0.51 & 4\\
Medium & 1650/1200 & -4.89 & 9.42 & 4\\
Medium & 1650/1350 & -4.04 & 7.93 & 4\\
Medium & 1800/1200 & -4.49 & 6.64 & 4\\
Medium & 1800/1350 & -2.21 & 5.00 & 4\\
Medium & 1800/1500 & -2.22 & 2.50 & 4\\
Medium & Gov/1500 & -1.80 & 2.09 & 4\\
Medium & Gov/1350 & -2.32 & 4.71 & 4\\
Medium & 1800/Gov & -0.80 & 0.41 & 4\\
Medium & 1650/Gov & -2.68 & 3.01 & 4\\
Medium & Gov/1200 & -3.27 & 6.36 & 4\\
Medium & 1800/1650 & -2.01 & 0.68 & 4\\
Large & 1200/1200 & -11.38 & 23.02 & 4\\
Large & 1500/1500 & -2.64 & 8.72 & 4\\
Large & 1800/1800 & -0.88 & -0.44 & 4\\
Large & 1980/1980 & -0.95 & -1.02 & 4\\
Large & 1650/1200 & -6.95 & 10.50 & 4\\
Large & 1650/1350 & -5.52 & 8.31 & 4\\
Large & 1800/1200 & -6.03 & 7.70 & 4\\
Large & 1800/1350 & -4.54 & 5.79 & 4\\
Large & 1800/1500 & -2.07 & 2.52 & 4\\
Large & Gov/1500 & -2.11 & 2.37 & 4\\
Large & Gov/1350 & -4.35 & 5.25 & 4\\
Large & 1800/Gov & -1.16 & -0.19 & 4\\
Large & 1650/Gov & -1.90 & 2.66 & 4\\
Large & Gov/1200 & -5.37 & 7.67 & 4\\
Large & 1800/1650 & -1.52 & 0.17 & 4\\
\bottomrule\end{longtable}}
\Needspace{10\baselineskip}
\subsection{Colocated synchronous: Nemotron, 32 GB200}
\noindent Four matched seeds per comparison.
{\small\setlength{\tabcolsep}{4pt}
\begin{longtable}{llrrl}
\toprule
Workload / split & Gen/train MHz & $\Delta q$ (\%) & Energy saving (\%) & $n$\\\midrule\endhead
Small & 1200/1200 & -12.99 & 7.49 & 4\\
Small & 1500/1500 & -8.59 & 5.78 & 4\\
Small & 1800/1800 & -3.01 & 2.95 & 4\\
Small & 1980/1980 & -0.84 & 0.43 & 4\\
Small & 1650/1200 & -5.00 & 6.61 & 4\\
Small & 1650/1350 & -5.14 & 4.85 & 4\\
Small & 1800/1200 & -2.53 & 6.16 & 4\\
Small & 1800/1350 & -3.27 & 4.24 & 4\\
Small & 1800/1500 & -3.52 & 4.07 & 4\\
Medium & 1200/1200 & -13.61 & 7.52 & 4\\
Medium & 1500/1500 & -4.68 & 7.58 & 4\\
Medium & 1800/1800 & -1.67 & 2.71 & 4\\
Medium & 1980/1980 & +0.32 & -0.19 & 4\\
Medium & 1650/1200 & -6.29 & 5.81 & 4\\
Medium & 1650/1350 & -5.08 & 6.44 & 4\\
Medium & 1800/1200 & -4.94 & 4.73 & 4\\
Medium & 1800/1350 & -3.91 & 4.50 & 4\\
Medium & 1800/1500 & -1.79 & 5.08 & 4\\
Large & 1200/1200 & -14.52 & 7.32 & 4\\
Large & 1500/1500 & -6.95 & 6.52 & 4\\
Large & 1800/1800 & -3.99 & 3.34 & 4\\
Large & 1980/1980 & -1.33 & -0.27 & 4\\
Large & 1650/1200 & -8.68 & 7.00 & 4\\
Large & 1650/1350 & -6.47 & 5.27 & 4\\
Large & 1800/1200 & -8.80 & 3.30 & 4\\
Large & 1800/1350 & -6.79 & 3.88 & 4\\
Large & 1800/1500 & -8.86 & 2.76 & 4\\
\bottomrule\end{longtable}}
\Needspace{10\baselineskip}
\subsection{Disaggregated synchronous: Qwen, H100 Gen8/Train32}
\noindent Four matched seeds per comparison.
{\small\setlength{\tabcolsep}{4pt}
\begin{longtable}{llrrl}
\toprule
Workload / split & Gen/train MHz & $\Delta q$ (\%) & Energy saving (\%) & $n$\\\midrule\endhead
Small & 1980/1980 & -1.36 & 11.25 & 4\\
Small & 1800/1500 & -1.56 & 16.19 & 4\\
Small & 1650/1500 & -3.12 & 17.40 & 4\\
Small & 1500/1500 & -4.88 & 18.44 & 4\\
Small & 1500/1350 & -5.68 & 21.59 & 4\\
Small & 1500/1200 & -5.95 & 23.38 & 4\\
Small & 1200/1500 & -13.09 & 18.57 & 4\\
Small & 1200/1350 & -12.56 & 22.25 & 4\\
Small & 1200/1200 & -12.87 & 24.16 & 4\\
Small & 810/810 & -34.62 & 8.19 & 4\\
Medium & 1980/1980 & -0.44 & 11.76 & 4\\
Medium & 1800/1350 & -1.36 & 21.27 & 4\\
Medium & 1650/1350 & -3.09 & 22.14 & 4\\
Medium & 1500/1500 & -4.78 & 19.53 & 4\\
Medium & 1500/1350 & -3.53 & 23.72 & 4\\
Medium & 1500/1200 & -5.24 & 25.44 & 4\\
Medium & 1200/1500 & -10.63 & 20.79 & 4\\
Medium & 1200/1350 & -10.71 & 24.24 & 4\\
Medium & 1200/1200 & -10.61 & 26.80 & 4\\
Medium & 810/810 & -30.71 & 15.10 & 4\\
Large & 1980/1980 & +0.58 & 13.23 & 4\\
Large & 1800/1200 & -0.82 & 24.73 & 4\\
Large & 1650/1200 & -1.45 & 25.82 & 4\\
Large & 1500/1500 & -1.41 & 21.74 & 4\\
Large & 1500/1350 & -2.79 & 25.01 & 4\\
Large & 1500/1200 & -0.36 & 28.23 & 4\\
Large & 1200/1500 & -4.05 & 24.72 & 4\\
Large & 1200/1350 & -3.84 & 28.17 & 4\\
Large & 1200/1200 & -4.45 & 30.26 & 4\\
\bottomrule\end{longtable}}
\Needspace{10\baselineskip}
\subsection{Disaggregated synchronous: Qwen, H100 Gen16/Train32}
\noindent Four matched seeds per comparison.
{\small\setlength{\tabcolsep}{4pt}
\begin{longtable}{llrrl}
\toprule
Workload / split & Gen/train MHz & $\Delta q$ (\%) & Energy saving (\%) & $n$\\\midrule\endhead
Small & 1980/1980 & -1.73 & 8.31 & 4\\
Small & 1800/1200 & -2.16 & 18.11 & 4\\
Small & 1650/1200 & -5.08 & 18.63 & 4\\
Small & 1500/1500 & -6.94 & 15.25 & 4\\
Small & 1500/1350 & -6.65 & 18.24 & 4\\
Small & 1500/1200 & -6.54 & 20.55 & 4\\
Small & 1200/1500 & -14.32 & 17.36 & 4\\
Small & 1200/1350 & -14.36 & 20.50 & 4\\
Small & 1200/1200 & -14.31 & 22.31 & 4\\
Small & 810/810 & -36.06 & 6.32 & 4\\
Medium & 1980/1980 & -0.74 & 9.00 & 4\\
Medium & 1800/1200 & -1.32 & 20.55 & 4\\
Medium & 1650/1200 & -3.31 & 21.88 & 4\\
Medium & 1500/1500 & -3.46 & 18.22 & 4\\
Medium & 1500/1350 & -4.58 & 21.10 & 4\\
Medium & 1500/1200 & -4.29 & 23.71 & 4\\
Medium & 1200/1500 & -9.57 & 20.73 & 4\\
Medium & 1200/1350 & -9.32 & 24.41 & 4\\
Medium & 1200/1200 & -10.65 & 26.12 & 4\\
Medium & 810/810 & -30.61 & 14.96 & 4\\
Large & 1980/1980 & -0.00 & 8.99 & 4\\
Large & 1800/1200 & -1.85 & 22.15 & 4\\
Large & 1650/1200 & -2.15 & 24.05 & 4\\
Large & 1500/1500 & -2.90 & 18.70 & 4\\
Large & 1500/1350 & -3.48 & 22.68 & 4\\
Large & 1500/1200 & -3.32 & 25.59 & 4\\
Large & 1200/1500 & -6.50 & 21.66 & 4\\
Large & 1200/1350 & -6.85 & 25.75 & 4\\
Large & 1200/1200 & -8.02 & 27.69 & 4\\
Large & 810/810 & -24.64 & 20.85 & 4\\
\bottomrule\end{longtable}}
\Needspace{10\baselineskip}
\subsection{Disaggregated synchronous: Qwen, H100 Gen8/Train24}
\noindent Four matched seeds per comparison.
{\small\setlength{\tabcolsep}{4pt}
\begin{longtable}{llrrl}
\toprule
Workload / split & Gen/train MHz & $\Delta q$ (\%) & Energy saving (\%) & $n$\\\midrule\endhead
Small & 1800/1650 & -1.84 & 12.17 & 4\\
Small & 1800/1500 & -1.86 & 15.63 & 4\\
Small & 1650/1650 & -2.26 & 14.33 & 4\\
Small & 1650/1500 & -3.53 & 16.30 & 4\\
Medium & 1800/1500 & -1.27 & 15.67 & 4\\
Medium & 1800/1350 & -1.14 & 20.00 & 4\\
Medium & 1650/1500 & -2.81 & 17.00 & 4\\
Medium & 1650/1350 & -3.52 & 20.63 & 4\\
Large & 1800/1500 & -0.96 & 16.38 & 4\\
Large & 1800/1425 & -2.08 & 18.03 & 4\\
Large & 1800/1350 & -2.06 & 20.16 & 4\\
Large & 1800/1200 & -3.16 & 22.05 & 4\\
Large & 1650/1350 & -3.36 & 21.14 & 4\\
Large & 1650/1200 & -2.23 & 24.21 & 4\\
\bottomrule\end{longtable}}
\Needspace{10\baselineskip}
\subsection{Disaggregated synchronous: Qwen, GB200}
\noindent Four matched seeds per comparison. Rows marked 5--15 cover updates 5--15 because at least one seed did not reach update 25 within the job's time limit. All other rows cover updates 5--25.
{\small\setlength{\tabcolsep}{4pt}
\begin{longtable}{llrrl}
\toprule
Workload / split & Gen/train MHz & $\Delta q$ (\%) & Energy saving (\%) & $n$\\\midrule\endhead
S 8/16 & 1800/1800 & -2.68 & 9.41 & 4\\
S 8/16 & 1400/1800 & -10.47 & 4.82 & 4\\
S 8/16 & 1800/1400 & -3.91 & 11.67 & 4\\
S 8/16 & 1400/1400 & -9.41 & 8.28 & 4\\
\addlinespace[2pt]
M 8/16 & 1800/1800 & -0.62 & 10.51 & 4\\
M 8/16 & 1400/1800 & -5.88 & 8.22 & 4\\
M 8/16 & 1800/1400 & +1.55 & 16.26 & 4\\
M 8/16 & 1400/1400 & -4.27 & 11.53 & 4\\
\addlinespace[2pt]
L 8/16; 5--15 & 1800/1800 & -3.98 & 9.18 & 4\\
L 8/16; 5--15 & 1400/1800 & -3.67 & 12.81 & 4\\
L 8/16; 5--15 & 1800/1400 & -7.06 & 10.25 & 4\\
L 8/16; 5--15 & 1900/1900 & +0.02 & 13.14 & 4\\
\addlinespace[2pt]
S 12/24 & 1800/1800 & -1.15 & 10.44 & 4\\
S 12/24 & 1400/1800 & -8.06 & 7.51 & 4\\
S 12/24 & 1800/1400 & -0.44 & 14.31 & 4\\
S 12/24 & 1400/1400 & -8.54 & 10.25 & 4\\
\addlinespace[2pt]
M 12/24 & 1800/1800 & -0.73 & 10.62 & 4\\
M 12/24 & 1400/1800 & -4.89 & 9.61 & 4\\
M 12/24 & 1800/1400 & -0.47 & 13.27 & 4\\
M 12/24 & 1400/1400 & -3.18 & 14.67 & 4\\
\addlinespace[2pt]
L 12/24 & 1800/1800 & +1.43 & 14.10 & 4\\
L 12/24 & 1400/1800 & +0.53 & 14.96 & 4\\
L 12/24 & 1800/1400 & +1.41 & 18.28 & 4\\
L 12/24; 5--15 & 1400/1400 & -1.49 & 17.64 & 4\\
\bottomrule\end{longtable}}
\Needspace{10\baselineskip}
\subsection{Disaggregated synchronous: Nemotron, H100 Gen16/Train64}
\noindent Four matched seeds per comparison.
{\small\setlength{\tabcolsep}{4pt}
\begin{longtable}{llrrl}
\toprule
Workload / split & Gen/train MHz & $\Delta q$ (\%) & Energy saving (\%) & $n$\\\midrule\endhead
Small & 1350/1350 & -2.44 & 26.29 & 4\\
Small & 1500/1500 & -1.46 & 20.96 & 4\\
Small & 1350/1500 & -1.31 & 23.75 & 4\\
Small & 1500/1350 & -2.07 & 23.81 & 4\\
Small & 1500/1650 & -3.74 & 17.20 & 4\\
Small & 1650/1500 & -3.15 & 18.30 & 4\\
Medium & 1350/1350 & -1.16 & 26.14 & 4\\
Medium & 1500/1500 & -2.52 & 18.92 & 4\\
Medium & 1350/1500 & -1.77 & 21.71 & 4\\
Medium & 1500/1350 & -2.48 & 23.33 & 4\\
Medium & 1500/1650 & -0.01 & 17.11 & 4\\
Medium & 1650/1500 & -1.83 & 18.20 & 4\\
Large & 1350/1350 & -3.95 & 23.94 & 4\\
Large & 1500/1500 & -1.96 & 17.84 & 4\\
Large & 1350/1500 & -4.10 & 18.51 & 4\\
Large & 1500/1350 & -4.45 & 22.02 & 4\\
Large & 1500/1650 & -1.02 & 14.20 & 4\\
Large & 1650/1500 & -2.49 & 16.28 & 4\\
\bottomrule\end{longtable}}
\Needspace{10\baselineskip}
\subsection{Disaggregated synchronous: Nemotron, H100 Gen16/Train32}
\noindent Four matched seeds per comparison. Rows marked 5--15 cover updates 5--15 because at least one seed did not reach update 25 within the job's time limit. All other rows cover updates 5--25.
{\small\setlength{\tabcolsep}{4pt}
\begin{longtable}{llrrl}
\toprule
Workload / split & Gen/train MHz & $\Delta q$ (\%) & Energy saving (\%) & $n$\\\midrule\endhead
Small & 1350/1350 & -1.73 & 24.98 & 4\\
Small & 1500/1500 & -2.72 & 17.89 & 4\\
Small & 1350/1500 & -1.20 & 21.53 & 4\\
Small & 1500/1350 & -3.26 & 21.20 & 4\\
Small & 1500/1650 & -1.90 & 15.50 & 4\\
Small & 1650/1500 & -2.64 & 16.01 & 4\\
Medium & 1350/1350 & -4.22 & 23.50 & 4\\
Medium & 1500/1500 & -3.31 & 17.08 & 4\\
Medium & 1350/1500 & -1.28 & 20.22 & 4\\
Medium & 1500/1350 & -2.20 & 22.26 & 4\\
Medium & 1500/1650 & -2.64 & 14.25 & 4\\
Medium & 1650/1500 & -1.46 & 16.63 & 4\\
Large; 5--15 & 1350/1350 & -0.06 & 25.20 & 4\\
Large; 5--15 & 1500/1500 & +1.17 & 18.87 & 4\\
Large; 5--15 & 1350/1500 & +1.66 & 20.57 & 4\\
Large; 5--15 & 1500/1350 & +0.99 & 24.31 & 4\\
Large; 5--15 & 1500/1650 & +0.24 & 14.36 & 4\\
Large; 5--15 & 1650/1500 & +1.06 & 17.76 & 4\\
\bottomrule\end{longtable}}
\Needspace{10\baselineskip}
\subsection{Disaggregated synchronous: Nemotron, GB200}
\noindent Four matched seeds per comparison. Rows marked 5--15 cover updates 5--15 because at least one seed did not reach update 25 within the job's time limit.
{\small\setlength{\tabcolsep}{4pt}
\begin{longtable}{llrrl}
\toprule
Workload / split & Gen/train MHz & $\Delta q$ (\%) & Energy saving (\%) & $n$\\\midrule\endhead
S 8/16; 5--15 & 1400/1400 & -4.99 & 13.31 & 4\\
S 8/16; 5--15 & 1900/1800 & +1.40 & 15.72 & 4\\
M 8/16; 5--15 & 1400/1400 & -0.88 & 16.27 & 4\\
M 8/16; 5--15 & 1800/1800 & -0.73 & 9.42 & 4\\
M 8/16; 5--15 & 1400/1800 & +1.98 & 16.08 & 4\\
M 8/16; 5--15 & 1800/1400 & -2.46 & 13.66 & 4\\
M 8/16; 5--15 & 1800/1900 & +0.64 & 10.55 & 4\\
\bottomrule\end{longtable}}

\Needspace{5\baselineskip}
\section{Asynchronous Headroom and Training Reward}
\label{sec:asyncinventory}
The tables report 67 comparisons over updates 5--25, each with four matched seeds and a native-governor reference over the same interval: 32 drained and 31 in-flight H100 comparisons, plus two Qwen comparisons of each mode on GB200. Drained and in-flight execution use policy ages 1 and 2, respectively. No shorter-window estimates or comparisons with fewer than four paired seeds are included. Incomplete or invalid comparisons are omitted, including large Nemotron in-flight at 16/32 H100s and large Qwen drained runs.

Energy saving is the reduction in whole-allocation GPU energy per valid training token, and $\Delta q$ is the change in valid training tokens per second. Splits give generation/training GPU counts, S/M/L denote workload sizes, and clock pairs give requested generation/training MHz. Each comparison retains its study's idle-control policy.

Reward differences are candidate minus governor in absolute reward units. Training reward is a diagnostic, not final validation accuracy.
\setcounter{table}{0}
\renewcommand{\thetable}{D\arabic{table}}
\renewcommand{\theHtable}{appendix.D.\arabic{table}}
\Needspace{7\baselineskip}
\stepcounter{table}
\noindent\textbf{Table~\thetable.} Qwen, H100, asynchronous drained; updates 5--25. Four matched seeds per row.\par\smallskip
{\footnotesize\setlength{\tabcolsep}{4pt}
\addtocounter{table}{-1}
\begin{longtable}{@{}llcrrr@{}}
\toprule
Split & Size & Gen/train MHz & Energy saving (\%) & $\Delta q$ (\%) & $\Delta$ reward\\\midrule\endfirsthead
\multicolumn{6}{l}{\textit{Qwen, H100, asynchronous drained; updates 5--25. (continued)}}\\
\toprule
Split & Size & Gen/train MHz & Energy saving (\%) & $\Delta q$ (\%) & $\Delta$ reward\\\midrule\endhead
\bottomrule\endfoot
8/24 & S & 1725/810 & +19.36 & -3.29 & -0.0046\\
8/24 & S & 1830/345 & +15.75 & -3.93 & -0.0023\\
8/24 & S & 1830/525 & +17.91 & -2.29 & -0.0017\\
8/24 & S & 1830/810 & +18.73 & -1.67 & +0.0015\\
8/24 & S & 1830/1200 & +17.53 & -0.84 & +0.0007\\
\addlinespace[3pt]
8/24 & M & 1725/810 & +20.95 & -2.66 & +0.0009\\
8/24 & M & 1830/345 & +17.28 & -2.06 & +0.0011\\
8/24 & M & 1830/525 & +19.10 & -1.42 & -0.0001\\
8/24 & M & 1830/810 & +19.89 & -1.37 & -0.0006\\
8/24 & M & 1830/1200 & +19.12 & -0.85 & +0.0004\\
\addlinespace[3pt]
16/32 & S & 1725/810 & +17.40 & -2.88 & +0.0028\\
16/32 & S & 1830/345 & +14.39 & -2.64 & +0.0045\\
16/32 & S & 1830/525 & +15.69 & -2.13 & +0.0020\\
16/32 & S & 1830/810 & +16.30 & -1.51 & +0.0035\\
16/32 & S & 1830/1200 & +15.33 & -1.09 & +0.0007\\
\addlinespace[3pt]
16/32 & M & 1725/810 & +18.50 & -2.26 & -0.0003\\
16/32 & M & 1830/345 & +8.42 & -15.72 & -0.0010\\
16/32 & M & 1830/525 & +16.32 & -1.73 & -0.0003\\
16/32 & M & 1830/810 & +16.74 & -2.38 & -0.0003\\
16/32 & M & 1830/1200 & +15.60 & -2.12 & +0.0000\\
\end{longtable}}

\Needspace{7\baselineskip}
\stepcounter{table}
\noindent\textbf{Table~\thetable.} Nemotron, H100, asynchronous drained; updates 5--25. Four matched seeds per row.\par\smallskip
{\footnotesize\setlength{\tabcolsep}{4pt}
\addtocounter{table}{-1}
\begin{longtable}{@{}llcrrr@{}}
\toprule
Split & Size & Gen/train MHz & Energy saving (\%) & $\Delta q$ (\%) & $\Delta$ reward\\\midrule\endfirsthead
\multicolumn{6}{l}{\textit{Nemotron, H100, asynchronous drained; updates 5--25. (continued)}}\\
\toprule
Split & Size & Gen/train MHz & Energy saving (\%) & $\Delta q$ (\%) & $\Delta$ reward\\\midrule\endhead
\bottomrule\endfoot
16/32 & S & 1725/810 & +19.07 & -1.68 & -0.0019\\
16/32 & S & 1830/810 & +17.83 & +0.72 & -0.0017\\
16/32 & S & 1830/1200 & +17.14 & +0.95 & +0.0000\\
16/32 & S & 1830/1500 & +9.11 & +0.14 & +0.0045\\
\addlinespace[3pt]
16/32 & M & 1830/1200 & +17.41 & -2.94 & -0.0025\\
\addlinespace[3pt]
16/64 & S & 1725/810 & +21.46 & -2.92 & +0.0000\\
16/64 & S & 1830/810 & +20.80 & -1.84 & -0.0007\\
16/64 & S & 1830/1200 & +20.16 & -2.12 & -0.0045\\
16/64 & S & 1830/1500 & +13.70 & -3.07 & +0.0028\\
\addlinespace[3pt]
16/64 & M & 1725/810 & +21.85 & -1.83 & +0.0015\\
16/64 & M & 1830/1200 & +21.13 & -0.26 & -0.0038\\
16/64 & M & 1830/1500 & +13.91 & +2.41 & -0.0025\\
\end{longtable}}

\Needspace{7\baselineskip}
\stepcounter{table}
\noindent\textbf{Table~\thetable.} Qwen, GB200, asynchronous drained; updates 5--25. Four matched seeds per row.\par\smallskip
{\footnotesize\setlength{\tabcolsep}{4pt}
\addtocounter{table}{-1}
\begin{longtable}{@{}llcrrr@{}}
\toprule
Split & Size & Gen/train MHz & Energy saving (\%) & $\Delta q$ (\%) & $\Delta$ reward\\\midrule\endfirsthead
\multicolumn{6}{l}{\textit{Qwen, GB200, asynchronous drained; updates 5--25. (continued)}}\\
\toprule
Split & Size & Gen/train MHz & Energy saving (\%) & $\Delta q$ (\%) & $\Delta$ reward\\\midrule\endhead
\bottomrule\endfoot
4/16 & M & 1900/1400 & +7.60 & -3.10 & +0.0006\\
4/16 & M & 1900/1700 & +7.84 & -1.90 & +0.0005\\
\end{longtable}}

\Needspace{7\baselineskip}
\stepcounter{table}
\noindent\textbf{Table~\thetable.} Qwen, H100, asynchronous in-flight; updates 5--25. Four matched seeds per row.\par\smallskip
{\footnotesize\setlength{\tabcolsep}{4pt}
\addtocounter{table}{-1}
\begin{longtable}{@{}llcrrr@{}}
\toprule
Split & Size & Gen/train MHz & Energy saving (\%) & $\Delta q$ (\%) & $\Delta$ reward\\\midrule\endfirsthead
\multicolumn{6}{l}{\textit{Qwen, H100, asynchronous in-flight; updates 5--25. (continued)}}\\
\toprule
Split & Size & Gen/train MHz & Energy saving (\%) & $\Delta q$ (\%) & $\Delta$ reward\\\midrule\endhead
\bottomrule\endfoot
8/24 & S & 1830/810 & +18.22 & -2.00 & +0.0060\\
8/24 & S & 1830/1200 & +17.06 & -0.80 & -0.0008\\
\addlinespace[3pt]
8/24 & M & 1725/1200 & +18.60 & -2.02 & +0.0013\\
8/24 & M & 1830/810 & +19.11 & -1.47 & -0.0007\\
8/24 & M & 1830/1200 & +17.68 & -0.89 & +0.0033\\
\addlinespace[3pt]
8/24 & L & 1725/1200 & +19.00 & -2.27 & -0.0002\\
8/24 & L & 1830/810 & +19.91 & -0.78 & -0.0009\\
8/24 & L & 1830/1200 & +17.93 & -0.93 & +0.0001\\
\addlinespace[3pt]
16/32 & S & 1725/1200 & +16.02 & -2.12 & +0.0112\\
16/32 & S & 1830/810 & +15.39 & -2.20 & +0.0086\\
16/32 & S & 1830/1200 & +14.49 & -0.96 & +0.0060\\
\addlinespace[3pt]
16/32 & M & 1725/1200 & +17.78 & -1.01 & +0.0009\\
16/32 & M & 1830/810 & +16.89 & -1.16 & +0.0014\\
16/32 & M & 1830/1200 & +16.07 & -0.24 & -0.0004\\
\addlinespace[3pt]
16/32 & L & 1725/1200 & +17.37 & -1.92 & -0.0011\\
16/32 & L & 1830/810 & +17.43 & -0.42 & -0.0003\\
16/32 & L & 1830/1200 & +16.36 & -0.09 & +0.0001\\
\end{longtable}}

\Needspace{7\baselineskip}
\stepcounter{table}
\noindent\textbf{Table~\thetable.} Nemotron, H100, asynchronous in-flight; updates 5--25. Four matched seeds per row.\par\smallskip
{\footnotesize\setlength{\tabcolsep}{4pt}
\addtocounter{table}{-1}
\begin{longtable}{@{}llcrrr@{}}
\toprule
Split & Size & Gen/train MHz & Energy saving (\%) & $\Delta q$ (\%) & $\Delta$ reward\\\midrule\endfirsthead
\multicolumn{6}{l}{\textit{Nemotron, H100, asynchronous in-flight; updates 5--25. (continued)}}\\
\toprule
Split & Size & Gen/train MHz & Energy saving (\%) & $\Delta q$ (\%) & $\Delta$ reward\\\midrule\endhead
\bottomrule\endfoot
16/32 & S & 1725/1200 & +17.67 & -1.78 & -0.0041\\
16/32 & S & 1830/810 & +10.05 & -13.26 & -0.0024\\
16/32 & S & 1830/1200 & +15.92 & -2.14 & -0.0058\\
\addlinespace[3pt]
16/32 & M & 1725/1200 & +18.12 & -4.05 & -0.0035\\
16/32 & M & 1830/810 & +13.00 & -15.68 & -0.0020\\
16/32 & M & 1830/1200 & +16.38 & -4.21 & +0.0012\\
\addlinespace[3pt]
16/64 & S & 1725/1200 & +21.69 & +1.48 & -0.0047\\
16/64 & S & 1830/810 & +18.64 & -5.18 & -0.0037\\
\addlinespace[3pt]
16/64 & M & 1725/1200 & +21.27 & -1.37 & +0.0052\\
16/64 & M & 1830/810 & +15.70 & -13.57 & -0.0093\\
16/64 & M & 1830/1200 & +20.83 & +0.11 & -0.0058\\
\addlinespace[3pt]
16/64 & L & 1725/1200 & +19.23 & -7.33 & +0.0009\\
16/64 & L & 1830/810 & +12.32 & -25.93 & -0.0009\\
16/64 & L & 1830/1200 & +18.52 & -7.35 & +0.0012\\
\end{longtable}}

\Needspace{7\baselineskip}
\stepcounter{table}
\noindent\textbf{Table~\thetable.} Qwen, GB200, asynchronous in-flight; updates 5--25. Four matched seeds per row.\par\smallskip
{\footnotesize\setlength{\tabcolsep}{4pt}
\addtocounter{table}{-1}
\begin{longtable}{@{}llcrrr@{}}
\toprule
Split & Size & Gen/train MHz & Energy saving (\%) & $\Delta q$ (\%) & $\Delta$ reward\\\midrule\endfirsthead
\multicolumn{6}{l}{\textit{Qwen, GB200, asynchronous in-flight; updates 5--25. (continued)}}\\
\toprule
Split & Size & Gen/train MHz & Energy saving (\%) & $\Delta q$ (\%) & $\Delta$ reward\\\midrule\endhead
\bottomrule\endfoot
4/16 & M & 1900/1400 & +8.91 & -1.24 & +0.0015\\
4/16 & M & 1900/1700 & +4.85 & +0.67 & +0.0004\\
\end{longtable}}

\Needspace{5\baselineskip}
\section{Workload-Normalized Controller Results}
\label{supp:normalized}
Update time and energy depend on the work each update receives. Response lengths and token counts differ between runs even with the same code and seed, because sampling, node timing, and the policy itself drift apart. These differences often reach 2--3\% of throughput, the same size as the clock effects we measure. A direct comparison with one governor run therefore cannot fully separate the cost of the clocks from a difference in work. We repeat each controller comparison against a governor model fitted to the work, using the controller run and two governor runs of the same configuration.

This model is fitted offline and is separate from the controller's online predictor. For each model and execution mode, we pool every update from both governor runs, starting at update 2, and fit update time and GPU energy by ordinary least squares on each update's workload features. Only governor updates enter the fit, so the model describes only what governor clocks do. The features follow what sets update time in each mode. In synchronous execution, generation is fitted on generated tokens $G$ and longest-response tokens $L$, training on valid training tokens $N$, and the remaining time is the governor median. With drained updates, Mistral-NeMo uses $L$ and $G$ and Qwen uses $G$ and $N$, since nearly every Qwen update reaches the 4096-token limit and $L$ barely varies. With in-flight updates, the fit uses the valid training tokens and longest response of the batch each update consumes, together with the previous update's valid training tokens, because in-flight work spills into the next update.

For each interval, a run's ratio is its measured total divided by the total the model predicts for the same updates. A ratio of one means the run matched the governor on its own work. The governor reference is the mean of the two governor ratios. Throughput change is the governor ratio divided by the controller ratio, minus one, and energy saving is one minus the controller ratio divided by the governor ratio. Ratios of sums weight each update by its length, as the totals in Table~\ref{tab:controller} do. All runs are scored on the same updates, and search, transition, and clock-switch updates all count.

Pooling two governor runs widens the range of work the model covers, so it predicts the controller's updates without extrapolating. It also reduces the weight of any one governor run, each of which has its own nodes and small slowdowns. Because one model scores both governors, the spread of their ratios shows how much variation remains after accounting for work. Over each full run, every governor lies within 0.7\% of the pooled model in time.

\begin{center}\small\setlength{\tabcolsep}{4pt}
\begin{tabular}{lrrrrrr}\toprule
 & \multicolumn{3}{c}{Whole run} & \multicolumn{3}{c}{After lock}\\\cmidrule(lr){2-4}\cmidrule(l){5-7}
Run & Updates & Energy saving & Throughput & Updates & Energy saving & Throughput\\\midrule
Qwen, colocated sync. & 2--107 & 7.1\% & $-$1.17\% & 78--107 & 8.3\% & $-$1.21\%\\
Qwen, disaggregated sync. & 2--93 & 16.4\% & $-$1.67\% & 62--93 & 22.0\% & $-$1.71\%\\
Mistral, colocated sync. & 2--265 & 3.1\% & $-$0.42\% & 173--265 & 2.5\% & +0.08\%\\
Mistral, disaggregated sync. & 2--315 & 9.3\% & $-$1.65\% & 60--315 & 9.7\% & $-$1.53\%\\
Qwen, async drained & 2--175 & 13.9\% & $-$2.19\% & 90--175 & 18.8\% & $-$0.95\%\\
Mistral, async drained & 2--300 & 11.7\% & $-$1.84\% & 191--300 & 18.7\% & $-$1.02\%\\
Qwen, async in-flight & 2--335 & 7.5\% & $-$1.25\% & 258--335 & 10.3\% & $-$1.57\%\\
Mistral, async in-flight & 2--560 & 6.6\% & $-$1.18\% & 314--560 & 7.7\% & $-$0.60\%\\
\bottomrule\end{tabular}\end{center}

Normalization keeps every energy saving within 2 points of Table~\ref{tab:controller}, and it moves most throughput changes by less than half a point. The exceptions are the Mistral-NeMo runs with throughput gains in Table~\ref{tab:controller}. After accounting for differences in work, both show small losses. Drained Qwen loses 2.19\% over its full run, slightly beyond the 2\% limit. 

\end{document}